\documentstyle[12pt]{article}

\input{epsf}

\def\R{\mbox{\bf R}}
\def\C{\mbox{\bf C}}
\def\H{\mbox{\bf H}}
\def\tenfour{\sigma_{10,4}}
\def\tenfive{\sigma_{10,5}}

\def\fiveten{\sigma_{5,10}}
\def\tensix{\sigma_{10,6}}

\def\fourfive{\sigma_{4,5}}
\def\fivefour{\sigma_{5,4}}
\def\sixfour{\sigma_{6,4}}
\def\foursix{\sigma_{4,6}}
\def\fivesix{\sigma_{5,6}}
\def\sixfive{\sigma_{6,5}}

\begin{document}
\begin{titlepage}
\begin{flushright}
PUPT-1714\\
ITEP-TH-33/97
\end{flushright}

\begin{center}
{\LARGE $ $ \\ $ $ \\ $ $ \\
BPS States and Minimal Surfaces}\\
\bigskip\bigskip\bigskip
{\Large A.Mikhailov}
\footnote{On leave from the
Institute of Theoretical and Experimental Physics,
117259, Bol. Cheremushkinskaya, 25, Moscow, Russia.}\\
\bigskip
Department of Physics, Princeton University\\
Princeton, NJ 08544, USA\\
\vskip 1cm
E-mail: andrei@puhep1.princeton.edu
\end{center}
\vskip 1in
\begin{abstract}
It was observed recently, that the low energy effective
action of the four-dimensional supersymmetric theories may be obtained
as a certain limit of M theory. From this point of view, the BPS states
correspond to the minimal area membranes ending on the M theory fivebrane.
We prove that for the configuration, corresponding to the $SU(2)$
Super Yang-Mills theory, the BPS spectrum is correctly reproduced,
and develop techniques for analyzing the BPS spectrum in more
general cases. We show that the type of the supermultiplet is related
to the topology of the membrane: disks correspond to hypermultiplets,
and cylinders to vector multiplets. We explain the relation between
minimal surfaces and geodesic lines, which shows that our description
of BPS states is closely related to one arising in Type II string 
compactification on Calabi-Yau threefolds.
\end{abstract}
\end{titlepage}

\section{Introduction.}
The exact solution of $N=2$ supersymmetric gauge theories
given by N.~Seiberg and E.~Witten in \cite{SW} played an
important role in the modern development of Quantum Field
Theory and String Theory.
It turned out, that the low energy effective action and some
massive states may be described in beautiful geometric language.
The moduli space of vacua corresponds to the moduli space of
certain 2-dimensional algebraic curves, and various physical
quantities are given by the very natural functions on it.

Recently
it turned out that this geometric language has a natural interpretation
in terms of string theory.
In particular, it was shown \cite{Witten}, that the low energy
effective action may be obtained as a special limit of
M-theory.

M-theory is the quantum theory whose low dimensional limit
is 11D supergravity \cite{Witten1,BFSS}. Its solitons are
5-branes and 2-branes.

The configuration considered in \cite{Witten} contains
a single M-theory 5-brane, with the worldvolume $\R^4\times\Sigma$,
embedded into the 11-dimensional space-time $\R^{10}\times S^1$.
Here $\R^4$ is parametrized by $x^0$, $x^1$, $x^2$ and $x^3$,
and $\Sigma$ is the surface in $\R^3\times S^1$. $\R^3$ has the coordinates
$x^4$, $x^5$ and $x^6$, and $S^1$ is parametrized by $x^{10}$.
The surface $\Sigma$ was given by the equation
\begin{equation}
F(x^6+ix^{10},x^4+ix^5)=0
\end{equation}

and it may be identified with the Seiberg-Witten curve of \cite{SW}.
The fact that $F$ is holomorphic means that the configuration preserves
$1/4$ of the 11-dimensional supersymmetries \cite{BBS,FS}. This gives
$N=2$ supersymmetry. Notice that for the flat 5-brane we would
have $N=4$ supersymmetry in $D=4$. From the point of view of string theory,
the curved 5-brane corresponds to a configuration of 5-branes
and 4-branes, which has half of the supersymmetry of the single
5-brane or 4-brane.
The asymptotical behaviour of $\Sigma$ fixes the function
$F$ up to several parameters, which are identified with the
moduli of $N=2$ vacua. The metric on the moduli space is
given by the kinetic energy of the motion of the 5-brane,
when we change these parameters. The BPS states correspond
to minimal area membranes ending on these 5-branes.

The masses of the BPS states in 4D $N=2$ theories are
given by the formula \cite{SW}:
\begin{equation}\label{BPS}
m_{n_e,n_m}=|n_e a+n_m a_D|
\end{equation}

where $n_e$ and $n_m$ are the electric and magnetic charges of the
BPS state, and $a(u)$, $a_D(u)$ are the functions of the order
parameters, given by the integral over the basic cycles in
$H_1(\Sigma,\mbox{\bf Z})$ of the meromorphic 1-form.
The set of allowed values of $n_e$ and $n_m$ carries important
information about the dynamics. Much work has been done in determining
which values are allowed \cite{FB,FB1}.

At some points $u$, $a/a_D$ becomes a rational number, and for
$n_e/n_m=-a_D/a$, formula (\ref{BPS}) implies that the mass becomes
zero. If the corresponding BPS states exist, this means
that at this point there is a singularity in the low energy
description of the theory. But in fact not all the values of $n_e$
and $n_m$ are allowed. In particular, it turns out that massless
states exist only when the corresponding homology cycles in $\Sigma$
actually contract to zero. This may be considered as one of
the indications that the curve $\Sigma$ itself, not just $a$, $a_D$,
has some physical meaning.

In this paper, we will study the BPS spectrum from the viewpoint
of M-theory. We will show that the correct BPS spectrum is reproduced
for $N=2$ SYM with the gauge group $SU(2)$, and develop techniques
for the study of BPS spectra in more general cases.

When this paper was in preparation, the related papers \cite{FS}
and \cite{YH} appeared.

\section{Membranes of minimal area.}
The projection of the 5-brane world-sheet to $S^1\times \R^3$
is the surface $\Sigma$, complex
in the complex structure $(x^6+ix^{10},x^4+ix^5)$.
This surface is non-compact (goes through infinity).
For the closed curve $\Gamma\subset \Sigma$, the winding number
$W[\Gamma]={1\over 2\pi}\int_{\Gamma}dx^{10}$ depends only on
the homology class $[\Gamma]\in H_1^{comp}(\Sigma,\mbox{\bf Z})$.
Here $H_1^{comp}(\Sigma)$ means $H_1$ of the curve $\Sigma$ with the points
at infinity omitted.
If $W[\Gamma]=0$, we can consider the membrane $M$ in $S^1\times \R^3$
whose boundary is $\Gamma$. Let $S[\Gamma]$ be the minimal
area of membranes
with the boundary $\Gamma$.
Suppose that for any homology class
$\gamma\in H_1^{comp}(\Sigma,\mbox{\bf Z})$,
the minimum $\mbox{min}_{\Gamma\in\gamma}S[\Gamma]$ is
realized on some curve $\Gamma_0(\gamma)$. The curve $\Gamma_0(\gamma)$
may have several connected components, then the corresponding
membrane can also have several connected components probably with different
topologies.

The area of the membrane is given by the integral
\begin{equation}
S[M]=\int_MdS=
\int_M\sqrt{\tenfour^2+\tenfive^2+\tensix^2+\fourfive^2+\foursix^2+
\fivesix^2}
\end{equation}

where $\sigma_{i,j}=dx^i\wedge dx^j(\sigma)$ are
the coordinates of the surface bivector. Since this bivector
is decomposable (it is a wedge product of two vectors, tangent to the
surface), it satisfies the
bilinear identity
\begin{equation}\label{bilinear}
\foursix\tenfive-\tensix\fourfive+\tenfour\sixfive=0
\end{equation}

we may rewrite $dS^2$ as ($s=x^6+ix^{10}$, $v=x^4+ix^5$):
\begin{equation}
dS^2=|ds\wedge dv|^2+(\tensix+\fivefour)^2
\end{equation}

This means that
\begin{equation}
S[\Gamma]\geq 
\left|\int_M ds\wedge dv\right|=\left|\int_{\Gamma}vds\right|
\end{equation}

The BPS states correspond to the membranes for which this
inequality is saturated.

This means that two conditions are satisfied:
\begin{equation}
\begin{array}{ll}
1) & \tensix+\fivefour=0;\\
2) & \mbox{Arg}(ds\wedge dv)=
\mbox{Arctan}\frac{\tenfour +\sixfive}{\sixfour+\fiveten}=\mbox{const}
\end{array}
\end{equation}

This is equivalent to the statement that the membrane is complex in the
complex structure

\begin{equation}
(\tilde{x^4}-i x^{10},\;\; x^6+i\tilde{x^5})
\end{equation}

where

\begin{equation}
(\tilde{x^4}+i\tilde{x^5})=e^{i\phi} (x^4+ix^5)
\end{equation}

and $\phi$ is a fixed angle.

Let us denote $\tilde{v}=\tilde{x}^4+i\tilde{x}^5$.
It is useful to introduce the complex coordinates $(x,y)$,
$x+y^*=s$, $x^*-y=\tilde{v}$. The 5-brane is holomorphic
in $(s,v)$, the 2-brane is holomorphic in $(x,y)$.

For the 5-brane $ds\wedge d\tilde{v}=0$, and
${4\over i}dx\wedge dy=d\sigma$ is the surface area.

For the 2-brane ${1\over i}ds\wedge d\tilde{v}=d\sigma$
is the surface area,
and $dx\wedge dy=0$.


Thus the question of classification of BPS states is reduced to the
classification of the curves in $C$ which are the lines of intersection
with the surface holomorphic in the other complex structure. Two surfaces,
holomorphic in two complex structures, generally intersect
at points. But there are exceptional cases when they may
intersect at curves. (In the latter case one can show that they
intersect at a right angle.)

Given {\em any} curve $\Gamma\in C$, we can in principle (by analytical
continuation) continue it to the complex surface $M$. The surface
$M$ will then be divided by $\Gamma$ into two parts, and we must
require that one of those parts is compact (does not go to
infinity).

\section{The simplest examples of membranes.}\label{examples}
The SW curve for $SU(2)$ is given by the equation
\begin{equation}
\cosh(s)-v^2+u=0
\end{equation}

Consider first the weak coupling region, that is, sufficiently large 
$|u|$ \cite{FB}. 
Topologically, the curve $\Sigma$ is the torus with two points
omitted. Its first homology group is generated by three cycles,
having winding numbers $1$, $1$ and $0$. They may be represented
by the equations $(x^5=0, x^4\in\left[\sqrt{u-1},\sqrt{u+1}\right])$,
$(x^5=0,x^4\in\left[-\sqrt{u+1},-\sqrt{u-1}\right])$ and
$(x^5=0, x^{10}=\pi)$ respectively.
For real $u>1$,
the projection of the curve onto the plane $(x^4,x^5,x^6)$ looks
like

\begin{center}
\leavevmode
\epsffile{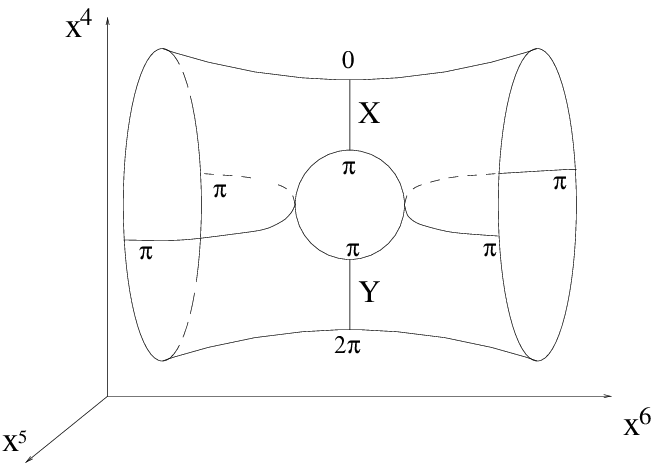}
\end{center}

where the values of $x^{10}$ are shown. Two line intervals
$X$ and $Y$ correspond to the points where the projection is
not one to one ($x^6=0$, so $x^{10}$ goes with $-x^{10}$.)

Consider the simplest examples of membranes:
\vskip 20pt
\epsfbox{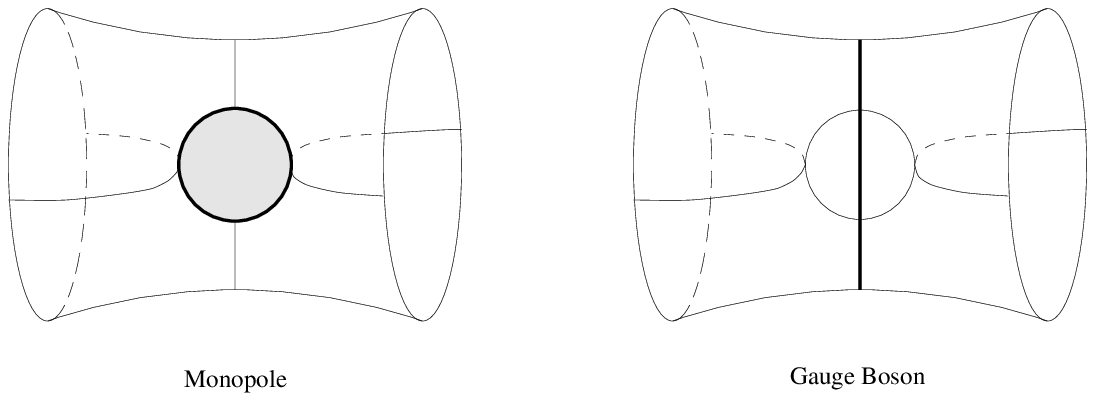}

A) \underline{Monopole.} $\gamma=(1,0)$, $m=|a_D|$.

The membrane has the topology of a disk. The equation of the
membrane is:

\begin{equation}\begin{array}{c}
x^5+ix^{10}=i\pi;\\
|x^4|\leq \sqrt{u-\cosh x^6}
\end{array}
\end{equation}

B) \underline{Vector Boson.} $\gamma=(0,2)$, $m=|2a|$.

This membrane has the topology of a cylinder, its
projection on the $(4,5,6)$ plane is just an interval. The equation of the
membrane is:

\begin{equation}
\begin{array}{c}
x^6+ix^5=0;\\
|x^4|\leq \sqrt{u+\cos x^{10}}
\end{array}
\end{equation}

The strong coupling regime corresponds to small $|u|$.
For real $u$ this means $u<1$.
In this case, the projection of the curve to $\R^3$ is two
surfaces, touching at two intervals.
These two intervals are

$$
x^5=0,\; x^4=\pm\sqrt{u+\cos x^{10}},\;\;
x^{10}\in [-\pi+\alpha,\pi-\alpha],
$$

and
$$
x^4=0, \; x^5=\pm\sqrt{-u-\cos x^{10}},\;\;
x^{10}\in [\pi-\alpha,\pi+\alpha]
$$

where $\alpha=\arccos u$

The compact homology has dimension three and is generated
by three cycles:
\begin{center}
\leavevmode
\epsfbox{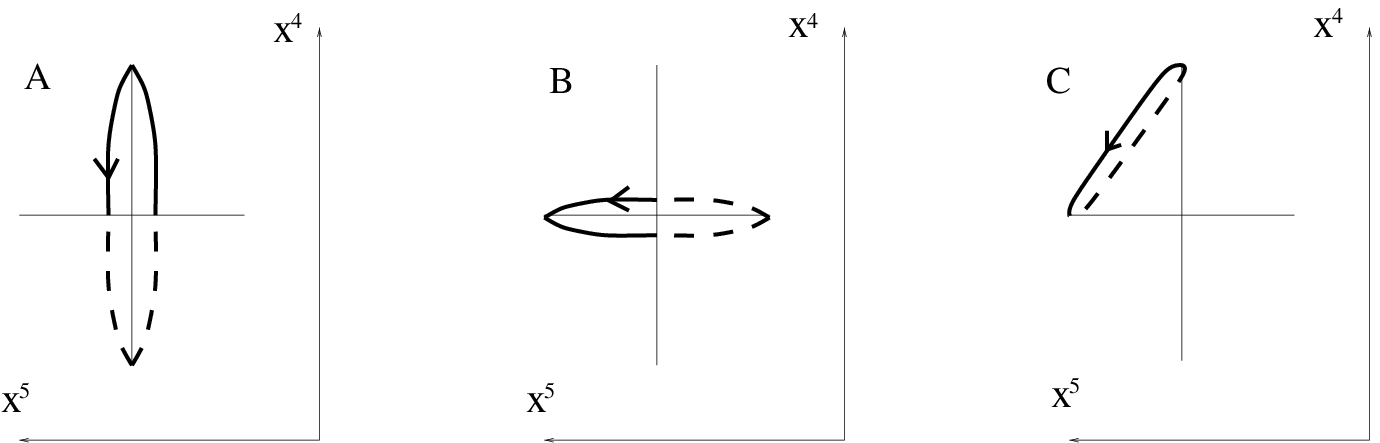}
\end{center}

(Shown is the view along the $x^6$-axis. We have two planes glued along the
cross.) Cycles A and B have winding number zero, and cycle C has
winding number 1.
To identify electric and magnetic cycles, we should follow
the deformation of these cycles at $u>1$
into the region $u<1$. In the process of deformation, it is
necessary to avoid the point $u=1$, since at that point the magnetic
cycle gets contracted, and we wouldn't be able to distinguish
between the cycle corresponding to the dyon and the cycle corresponding
to the vector boson. Thus, we have to turn on the imaginary part
of $u$. Since it is not very easy to draw the pictures at complex $u$,
we will consider the periods of $vds$ instead.
At real $u>1$,

\begin{equation}\begin{array}{c}\label{passing1}
a={i\sqrt{2}} \int\limits_0^{2\pi}\sqrt{u+\cos x^{10}}dx^{10}=
i\left(4-{u-1\over 2}\log{u-1\over 2}+\ldots\right)\\
a_D=\sqrt{2} \int\limits_{\cosh x^6<u}\sqrt{u-\cosh x^6}dx^6=
2\pi\left({u-1\over 2}+\ldots\right)
\end{array}
\end{equation}

There is the sign ambiguity in the magnetic
quantum number corresponding to the states at $u<1$, related to the
possibility to continue $a$ and $a_D$ from $u>1$ by going
above or below the point $u=1$ \cite{FB}. Let us assume that
we pass this point in the upper halfplane. At $u<1$, the period
of the differential over the A cycle is imaginary, and the period over
the B cycle is real. Comparing this to (\ref{passing1}), we
see that A corresponds to the dyon $(n_e,n_m)=(2,-1)$ and B corresponds
to the monopole $(n_e,n_m)=(0,1)$.
When $u\to 1$, $\alpha\to 0$, the horizontal interval shrinks, and
the monopole becomes massless.

In the case of real $u$, $|u|<1$,
we can write explicit expressions for these two disks:

A.) \underline{Dyon.} $x^6+ix^5=0$,
$|x^4|\leq\sqrt{u+\cos x^{10}}$.

B.) \underline{Monopole.} $x^6-ix^4=0$, $|x^5|\leq\sqrt{-u-\cos x^{10}}$.

The vector boson would correspond to
C--[--C], where --C means C transformed by $x^{4,5}\to-x^{4,5}$.
The corresponding membrane would have two boundaries,
the topology of a cylinder. But it turns out, that such a holomorphic
membrane does not exist. This follows from the arguments 
of the next section.

Notice that A+B=C--[--C] and intuitively it is probably clear
from the picture that the minimum is realized on the union of two
disks with boundaries A and B, not on the cylinder.

\section{Complex $u$.}\label{Complexu}

In this section we consider membranes for complex values of
the order parameter $u$. Suppose that we have constructed the
holomorphic membrane, corresponding to the given homology
cycle in $\Sigma_u$ for the given value of $u$. We will consider
what happens to this holomorphic membrane, when we start changing $u$.
We will show that when $a(u)/a_D(u)$ is not real,
the membrane changes smoothly. On the other hand, when it is real,
that is, when $u$ crosses the curve of marginal stability, there is the
possibility that the membrane decays into two membranes.

This result will enable us to construct the membranes for
general $u$ starting from real $u$, and
to construct membranes
corresponding to dyons applying the monodromy transformations
to the membrane corresponding to the monopole.

We will also show, that the membranes with the topology of a
disk have moduli space $\R^3$ (spatial 
translations), and the moduli space of the cylinder is $\R^3\times I$
(spatial translations plus one extra modulus, with the topology
of an interval).  

The membrane with the topology of a disk can be represented by
the holomorphic map
\begin{equation}
M: D\rightarrow \C^2
\end{equation}

from the disk $|z|<1$ to $\C^2$, such that the
image of the boundary of the disk, $z=e^{i\varphi}$,
lies on the surface of the 5-brane.
When we vary $u$, the surface of the 5-brane varies, and the variation may
be  thought of as the section of the normal bundle of $\Sigma$:
\begin{equation}\nonumber
\xi\in H^0(\Sigma,C^2/{\cal T}\Sigma)
\end{equation}

Consider the restriction of $\xi$ to the boundary of the membrane.
We may think
of this restriction as the $\C^2$-valued function on the circle,
modulo those functions whose values are tangent to the 5-brane.
Let us prove that we can always find a representative which can be
holomorphically continued inside the disk, to $|z|<1$. This representative
gives us the deformation of the surface of the membrane.

Let $(x,y)$ be the coordinates in $\C^2$ in which the membrane is
holomorphic, $x=(s-\tilde{v}^*)/2$, $y=(s^*+\tilde{v})/2$.
Let us denote $X_1=(\dot{x},\dot{y})$ the vector tangent to the boundary
of the membrane, $M_*{\partial\over\partial\varphi}$.
At any $\varphi$, this vector is tangent to the 5-brane, since
$\partial M \subset\Sigma$. Another vector tangent to the 5-brane is
$I(\dot{x},\dot{y})$, where $I$ is the operator of multiplication
by $i$ in the complex structure $(s,v)$. Explicitly
$X_2=(i\dot{y}^*,-i\dot{x}^*)$.

The deformation vector $\xi$, restricted to the boundary,
$$\xi(\varphi)=(\xi^x(\varphi),\xi^y(\varphi))$$
can be decomposed
into the two parts one of which can be holomorphically
continued inside the disk and the other outside:
\begin{equation}
\xi(\varphi)=\xi_{\geq 0}(\varphi)+\xi_{<0}(\varphi)
\end{equation}

Here $\xi_{\geq 0}$ and $\xi_{<0}$ may be defined 
in terms of Fourier series.
Given $$\xi(\varphi)=\sum\limits_{n=-\infty}^{\infty}
\xi_n e^{in\varphi}$$

we define 
$$\xi_{\geq 0}(\varphi)=\sum\limits_{n=0}^{\infty}
\xi_n e^{in\varphi}$$

and
$$\xi_{<0}(\varphi)=\sum\limits_{n=-\infty}^{-1}\xi_n e^{in\varphi}$$

We want to show that there exist 
real functions $a(\varphi)$, $b(\varphi)$ such that
\begin{equation}\label{negpart}
\xi_{<0}=[aX_1+bX_2]_{<0}
\end{equation}

Then  the vector-function $\tilde{\xi}(\varphi)=\xi-aX_1-bX_2$ can
be analytically continued inside the disk, and the map
\begin{equation}
M_{\epsilon}(z)=M(z)+\tilde{\xi}(z)
\end{equation}

will give the deformed membrane.

Let us first find $b$. Given two equations,
\begin{equation}\label{twoequations}
\begin{array}{c}
{[}a\dot{x}+ib\dot{y}^*{]}_{<0}=\xi^x_{<0}\\
{[}a\dot{y}-ib\dot{x}^*{]}_{<0}=\xi^y_{<0}
\end{array}
\end{equation}

we have
\begin{equation}
\left(\dot{y}\left[a\dot{x}+ib\dot{y}^*\right]_{<0}-
\dot{x}\left[a\dot{y}-ib\dot{x}^*\right]\right)_{\leq 0}=
(\dot{y}\xi^x_{<0}-\dot{x}\xi^y_{<0})_{\leq 0}
\end{equation}

Notice that $\dot{x}_{\leq 0}=0$.
Thus, for any function $\psi(\varphi)$, we get
${[}\psi_{<0}\dot{x}{]}_{\leq 0}=
 {[}\psi\dot{x}{]}_{\leq 0}$.
Taking this into account, we get:
\begin{equation}\label{b}
i{[}b(|\dot{x}|^2+|\dot{y}|^2){]}_{\leq 0}=
{[}\dot{y}\xi^x_{<0}-\dot{x}\xi^y_{<0}{]}_{\leq 0}
\end{equation}

Is this equation solvable for $b$? The obvious solution:
\begin{equation}\label{bsol}
ib=\frac{{[}\dot{y}\xi^x_{<0}-\dot{x}\xi^y_{<0}{]}_{\leq 0}-
{[}h.c.{]}_{>0}}
{|\dot{x}|^2+|\dot{y}|^2}
\end{equation}

is valid modulo two potential problems, which we will discuss later.
But if it works, then
$a(\varphi)$ can be found from either one of the two equations
\footnote{Written in terms of the Fourier coefficients, this is the
triangular system of equations, and always has a solution.}
(\ref{twoequations}). (Actually, $a_{-1}$, $a_0$ and $a_1=a_{-1}^*$ are
not determined from that equation --- this corresponds to the
$SL(2,\R)$ symmetry of the disk.)

The first problem is that the zero mode of
the LHS is purely imaginary:
$$
\mbox{\rm Re}\left(i{[}b(|\dot{x}|^2+|\dot{y}|^2){]}_0\right)=0
$$

So we have to prove that the zero mode of the RHS is also imaginary:
\begin{equation}
{[}\dot{y}\xi^x_{<0}-\dot{x}\xi^y_{<0}{]}_0=
\int_{\partial M} \xi^x dy-\xi^y dx\in i\R
\end{equation}

 At least locally, we may introduce in $\C^2$
the coordinate system $(s,u)$. The surfaces
$u=const$ are the curves $\Sigma_u$, and there is a function
$\tilde{v}(s,u)$ such that the integral of $\tilde{v}ds$
over the certain cycle in $\Sigma_u$ is purely imaginary
(that is how we defined $\tilde{v}$ in the first section).
Now

\begin{equation}\label{strip}
\begin{array}{c}
\int_{\partial M} \xi^x dy-\xi^y dx=
{1\over 4}\int_{\partial M}\left(
(\delta s-\delta\tilde{v}^*)d(s^*+\tilde{v})
-(\delta s^*+\delta\tilde{v})d(s-\tilde{v}^*)
\right)=\\=
\int_S ds\wedge ds^*+d\tilde{v}\wedge d\tilde{v}^*-
2\mbox{\rm Re}(d\tilde{v}\wedge ds)
\end{array}
\end{equation}

where $S$ is the strip between the membrane boundaries in $\Sigma_u$ and
$\Sigma_{u+\delta u}$. It is clear that the first two terms are imaginary.
The third one by the Stokes theorem is the real part of the
difference between the
integrals of  $\tilde{v}ds$ over the cycles in $\Sigma_u$ and
$\Sigma_{u+\delta u}$, and it is zero as follows from the previous
paragraph.

The second problem: it may happen that $|\dot{x}|^2+|\dot{y}|^2=0$ at
some point $\varphi_0$ on the boundary.
This means that the boundary of the membrane develops a cusp at this
point: the direction of the velocity vector on the left of $\varphi_0$ does
not coincide with the direction of the velocity vector on the right.
But this cannot happen with the boundary of a holomorphic  minimal
surface, unless the boundary touches itself. 

Indeed, suppose that
the boundary does not touch itself, and yet develops a cusp.
Without loss of generality, we may consider the case when the cusp
is at $z=-1$. Then, near the cusp, the equation of the curve may
be approximated as ($\zeta=z+1$): 
\begin{equation}\label{cuspapprox}
\begin{array}{c}
x=a\zeta^{\alpha}(1+o(1))\\
y=b\zeta^{\beta}(1+o(1))
\end{array}
\end{equation}

with $a$, $b$, $\alpha>1$, $\beta>1$ -- some parameters. 
The vectors, tangent to $\Sigma$ near the cusp of the membrane,
are:
\begin{equation}
\begin{array}{c}
(\dot{x},\dot{y})=i(a\alpha\zeta^{\alpha-1},b\beta\zeta^{\beta-1})= 
(A\zeta^{\alpha-1},B\zeta^{\beta-1})\\
(i\dot{y}^*,-i\dot{x}^*)=(iB^*(\zeta^*)^{\beta-1},
-iA^*(\zeta^*)^{\alpha-1})
\end{array}
\end{equation}

The surface element of $\Sigma$ (normalized to unit surface) has 
the $xy^*$ coordinates 
\begin{equation}
\sigma^{xy^*}=i\frac{A^2\zeta^{2\alpha-2}-
(B^*)^2(\zeta^*)^{2\beta-2}}{|A|^2|\zeta|^{2\alpha-2}+
|B|^2|\zeta|^{2\beta-2}}
\end{equation}

It follows from this expression, that $\sigma^{xy^*}$ is irregular at 
$\zeta=0$, that is the curve $\Sigma$ necessarily has singularity
at the point where the boundary of the membrane develops cusp.
For regular $\Sigma$, the cusps are impossible, except for the case
when the boundary of the membrane touches itself: in this case,
we cannot approximate the cusp by the simple equation (\ref{cuspapprox}).

In the case that the boundary touches itself,
the membrane is the union of two membranes. Their boundaries represent
different homology classes of $\Sigma$, and yet they are holomorphic
in the same complex structure. This means that 
the phases of the two periods of $vds$ coincide, 
that is we are on the curve of
marginal stability.

The conclusion is that to the deformation of the curve $\Sigma$
corresponds the
deformation of the membrane, except for possible decays
when we intersect the curve of marginal stability.

This consideration do not apply
to the membrane corresponding to the gauge
boson, since it has the topology of a cylinder.
For the cylinder, we will use somewhat different approach.
Introduce the coordinate $z=\tau+i\varphi$, so that the boundaries
are at $\tau=-t$ and $\tau=t$.
The deformation of the boundary of the cylinder may be
represented as:
\begin{equation}\label{cyl}
\begin{array}{c}
\xi^x=\alpha\frac{\overline{\partial y}}{w}+i\beta\partial x \\
\xi^y=-\alpha\frac{\overline{\partial x}}{w}+i\beta\partial y
\end{array}
\end{equation}

where we denoted
\begin{equation}
w=|\dot{x}|^2+|\dot{y}|^2=|\partial x|^2+|\partial y|^2
\end{equation}

and $\alpha=\alpha(\varphi)$ and $\beta=\beta(\varphi)$ are some
functions. The representation (\ref{cyl}) corresponds to
decomposition of $\xi$ in the basis $X_1$, $X_2$, $iX_1$, $iX_2$,
where $X_1=(\dot{x},\dot{y})$ and $X^2=IX^1$. 
For real $\alpha$ and $\beta$, $\xi$ is parallel
to $\Sigma$ (notice that on the boundary $\dot{x}=i\partial x$).
Thus, the element of the normal bundle ${\cal N}\Sigma$,
corresponding to $\xi$, depends only on $\mbox{Im}\alpha$ and
$\mbox{Im}\beta$.
Given the imaginary part
of the function $\alpha$ on the boundary,
we can determine the holomorphic
continuation inside the cylinder unambiguously.
The only requirement is that
\begin{equation}
\int d\varphi\; \mbox{Im}\;\alpha(t+i\varphi)=
\int d\varphi\; \mbox{Im}\;\alpha(-t+i\varphi)
\end{equation}

and this follows from (\ref{cyl}) and
(\ref{strip}). Let us also continue the function
$\beta$ inside the cylinder, not necessarily as a holomorphic 
function (the ambiguities in the continuation of $\beta$
correspond to the possible reparametrizations of the cylinder).
Now, the formulae (\ref{cyl}) determine some vector $\xi$ on the cylinder,
and the deformed surface
\begin{equation}
\begin{array}{c}
x'(z,\bar{z})=x(z)+\xi^x(z,\bar{z})\\
y'(z,\bar{z})=y(z)+\xi^y(z,\bar{z})
\end{array}
\end{equation}

turns out to be holomorphic in the complex structure $(x,y)$, although
$x'$ and $y'$ are not holomorphic functions of $z$. Indeed, the necessary
condition for the surface $(x(z,\bar{z}),y(z,\bar{z}))$ to be holomorphic
is
\begin{equation}
\{ x,y \}=\partial x\bar{\partial}y-
\bar{\partial}x \partial y=0
\end{equation}

and one can see that this condition is satisfied for
$(x+\xi^x,y+\xi^y)$ with $\xi^x$ and $\xi^y$ as in 
(\ref{cyl}).
It is possible to
introduce a new variable $z$ so that $x'$ and $y'$ are holomorphic
functions of $z$. We should not have expected that the deformed
membrane is given in parametric form by $x(z)$, $y(z)$ with the
same $z$ as the initial membrane. Indeed, this would mean that the deformed
membrane is isomorphic to the initial as complex manifold. But it is
generally not true, since the cylinder has a modulus, the length of
the cylinder.

Notice that the cylinder has nontrivial moduli space. Indeed, we may
take $\beta=0$, and $\alpha$ a real constant. This corresponds to sliding the
boundary of the cylinder along $\Sigma$, with the normal velocity
at point $\varphi$ equal
$$
{\alpha\over\sqrt{|\dot{x}|^2+|\dot{y}|^2}} 
$$

where $\dot{x}$ and $\dot{y}$ denote the derivative of $x$ and $y$ 
with respect to $\varphi$. 
In string theory language,
this zero mode corresponds to the motion of the string connecting two
fourbranes, in the direction $x^6$. This motion stops when the boundary
of the cylinder touches itself: 
\begin{center}
\leavevmode
\epsffile{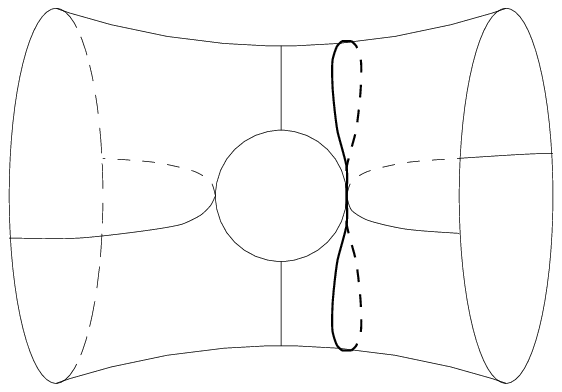}
\end{center}
\vskip -35pt

At this point, the length of the cylinder
goes to zero --- the cylinder degenerates into the disk.
When we map a very short cylinder into the four-dimensional
space, so that the image is of finite size, the derivatives $\dot{x}$
and $\dot{y}$ are typically very large, except for the small region
near the point where the boundary touches itself, and expressions 
(\ref{cyl}) become ill defined.  
Thus, the moduli space of the cylinder
has the topology of an interval.

We will generalize the construction of the bosonic zero modes
to the surfaces of more complicated topology in 
Section \ref{Deformations}.

Now we can apply the considerations of \cite{FB} to find the BPS spectrum.
First let us prove that the membrane corresponding to the dyon
exists ($u>1$). Notice that the curve $\Sigma_{-u}$ may be obtained from the
curve $\Sigma_u$ by the change of variables:
\begin{equation}\label{Z2}
\begin{array}{l}
y^6=x^6,\;\; y^{10}=x^{10}+\pi,\\
y^4=-x^5,\;\; y^5=x^4
\end{array}
\end{equation}

-- this is the manifestation of the $Z_2$ symmetry of the moduli space of
vacua.

Now the simplest membrane, with the topology of a disk, whose  
boundary is on $\Sigma_{-u}$,
\begin{equation}
\begin{array}{c}
y^5+iy^{10}=i\pi;\\
y^4\leq\sqrt{u-\mbox{cosh} y^6}
\end{array}
\end{equation}

goes to the membrane corresponding to the dyon when we smoothly
change $\Sigma_{-u}$ to $\Sigma_u$:
\vskip 20pt
\epsfbox{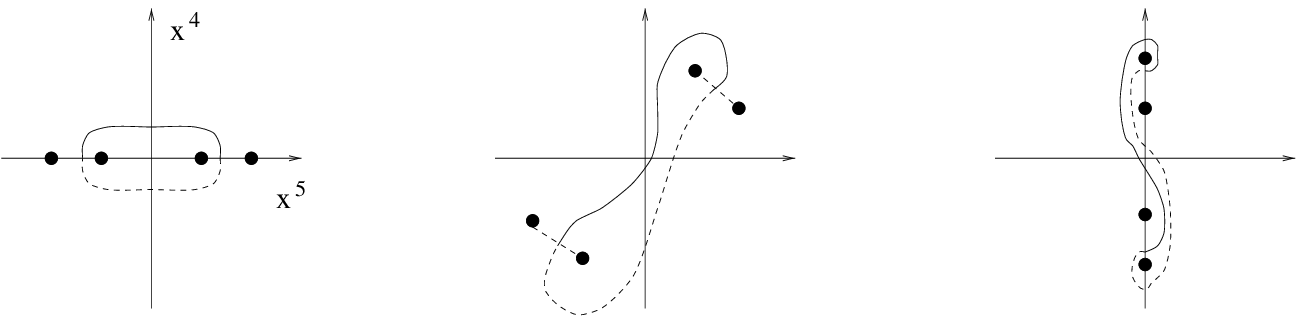}

The boundary of the disk corresponding to the dyon has the following 
shape:
\begin{center}
\leavevmode
\epsffile{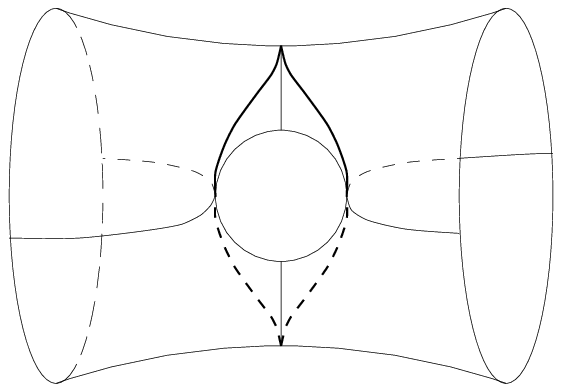}
\end{center}
\vskip -35pt

The membranes corresponding to the states $(n_e,n_m)=(2m,1)$ can
be constructed from monopole and dyon by moving $u$ around the
circle of the large radius.

Membranes with magnetic charge greater than one do not
exist, because if they existed, then  
we would be able to move $u$ to certain point on the curve
of marginal stability in
such a way, that the corresponding state becomes massless
--- see \cite{FB} for the details.
This would mean that the membrane of zero area has boundary
representing the nonzero homology class in $\Sigma$.

For the strong coupling regime, we have explicitly constructed the
membranes corresponding to the monopole and the dyon, and the
considerations from \cite{FB} show that no other states exist.
The proof goes as follows.
First, it is possible to prove that the existence of $(n_e,n_m)$
implies the existence of $(n_e+2n_m,- n_e - n_m)$. (Given $(n_e,n_m)$,
we know the state with the same mass exists at $-u$, because of the
$Z_2$ symmetry (\ref{Z2}), and deforming it from $-u$ to $u$
we get $(n_e+2n_m,-n_e-n_m)$.) But one of the states $(n_e,n_m)$ or
$(n_e+2n_m,-n_e-n_m)$ becomes massless at some point on the curve of
marginal stability. So, the only possibilities are $(0,\pm 1)$ and
$\pm (2,-1)$ -- the monopole and the dyon.

We have not proven that the membranes with more complicated
topology do not exist. If they exist, then their zero modes
may be described by the general construction, which we will
discuss in Section \ref{Deformations}.

\section{Matter in mixed representation.}

Besides the pure SYM, the theories with matter hypermultiplets were
described in \cite{Witten} in the M-theory language.
The configuration of branes corresponding to
the theories with hypermultiplets in fundamental representation
was constructed. Also, in the case when the gauge group
is $SU(n)^k$, the hypermultiplets in the mixed representation
(adjoint when $k=1$) were described.

In the case of mixed representation, the string theory fivebrane
is represented as a ``spike'' on the curve $\Sigma$, of the
form
\begin{equation}
v={m\over s-s_0} +\ldots
\end{equation}

where dots represent terms regular at $s=s_0$. The constant
$m$ is the mass of the matter hypermultiplet.
The membrane corresponding to this hypermultiplet has
the topology of a disk, the boundary of the disk being the small
circle around $s=s_0$. To prove the existence of such a membrane, 
we use the deformation argument from the previous section.

Let us deform the neighborhood of $s=s_0$
to the neighborhood of the point $s=s_0$ on the curve given
by $v={m\over s-s_0}$ (without any regular terms). We have to require
that $m$ is sufficiently small, so that the membrane is located
in the region around $s=s_0$, which can be deformed to the
corresponding region of the curve $v={m\over s-s_0}$.
Then the membrane is given by the equation

\begin{equation}\label{hypers}
\begin{array}{c}
v-(s^*-s_0^*)=0\\
|s-s_0|^2\leq m
\end{array}
\end{equation}

Since the membrane has the topology of a disk, it
corresponds to the hypermultiplet.
The masses of these states are just $|m|$.

\section{Matter in fundamental representation.}

To describe $N=2$ supersymmetric QCD with $N_f$ flavors,
it is necessary to introduce $N_f$ 6-branes.
The 6-brane in M-theory is described by replacing $\R^3\times S^1$ with
the Taub-NUT space. It may be obtained as the Hyper-Kahler reduction
of the flat quaternionic space $\H\times\H$
\cite{GR}. Introduce the quaternionic coordinates
$(q,w)$, $q=y+zj$, $w=u+vj$. This space is endowed with the flat
quaternion-valued symplectic form:
\begin{equation}
-{1\over 2} dq\wedge d\bar{q}-{1\over 2} dw\wedge d\bar{w}
=i\omega_I+j\omega_J+k\omega_K
\end{equation}

The moment map for the $\R$ action
\begin{equation}\label{Raction}
\begin{array}{c}
q\to qe^{it}=e^{it}y+e^{-it}zj\\
w\to w+t=(u+t)+vj
\end{array}
\end{equation}

takes values in imaginary quaternions:

\begin{equation}
\mu={1\over 2}qi\bar{q}-{1\over 2}(w-\bar{w})=i\mu_R+\mu_C j
\end{equation}

where
\begin{equation}
\begin{array}{c}
\mu_R={1\over 2}
(|y|^2-|z|^2-2 \mbox{Im}u)\\
\mu_C=izy+v
\end{array}
\end{equation}

The Taub-NUT space is obtained by first putting $\mu_R=\mu_C=0$
and then dividing by the action of $\R$. The space $\H\times\H$ has
three natural complex structures, which become complex structures
on the reduced space, as explained in Appendix A.
 But to describe the
brane configuration, it is more convenient to fix $\mu_C$ and then
divide by the action of $G_C=\C$ \cite{Witten}. $\mu_C=e$ gives
$zy=e-iv$, $\C$ acts as $y\to e^{it}y$, $z\to e^{-it}z$, $u\to u+t$.
To complete the reduction, we
introduce the $G_C$-invariant functions $Z=ze^{iu}$, $Y=ye^{-iu}$ on
the constraint manifold, and get the $\mbox{dim}_C=2$ manifold
\begin{equation}
\{(Z,Y,v)|ZY=e+iv\}
\end{equation}

In this formalism, one of the three complex structures of Taub-NUT is
more manifest then the other two. The surface $\Sigma$, representing
the 5-brane, is holomorphic in this complex structure.

BPS states are described as the membranes of minimal area.
The relation between the area of the surface in the Kahler
manifold and the integral of the holomorphic 2-form is a particular
case of the Wirtinger inequality \cite{Mumford,Lawson}.
It says, that the volume form
on the submanifold of the Kahler manifold is greater or equal to the
restriction of the appropriate power of the Kahler form.
The equality is when the submanifold is complex.
Let us consider the special case of the four-dimensional hyper-Kahlerian
manifold $X$. 
It has three complex structures, $I$, $J$ and $K$. For each
complex structure, consider the corresponding Kahler form:
\begin{equation}
\omega_{I_a}(\xi,\eta)=(\xi,I_a\eta)
\end{equation}
where $(,)$ is the (real) scalar product. Operators $I$, $J$ and $K$
generate the algebra $su(2)$.
The space of  bivectors $\Lambda^2TX$ decomposes into
the direct sum of one vector and three scalar representations of 
this $su(2)$. Let us denote the corresponding projectors $P_0$ and $P_1$.
For the decomposable bivector of the form $\xi\wedge\eta$, we have an
analogue of the bilinear identity (\ref{bilinear}):
\begin{equation}
|P_0(\xi\wedge\eta)|=|P_1(\xi\wedge\eta)| 
\end{equation}

where $|b|$ means the area of the bivector $b$. 
Indeed, this is the unique $su(2)$-invariant
condition which becomes (\ref{bilinear}) for the flat space.
Thus, for the area of the
surface element we have:
\begin{equation}
\begin{array}{c}
|\xi\wedge\eta|^2=2|P_1(\xi\wedge\eta)|^2=
-{1\over 4}\left(\xi\wedge\eta, 
\sum\limits_{a=1}^3 I_a^2 (\xi\wedge\eta)\right)
=\\=
{3\over 2}|\xi\wedge\eta|^2-{1\over 2}\left(\xi\wedge\eta,
\sum\limits_{a=1}^3 I_a\xi\wedge I_a\eta\right)
=\\=
{3\over 2} |\xi\wedge\eta|^2+{1\over 2}(\xi,I_a\eta)(\eta,I_a\xi)
\end{array}
\end{equation}

or 
\begin{equation}\label{ds2}
|\xi\wedge\eta|^2=\sum\limits_{a=1}^3\left[
\omega_a(\xi,\eta)\right]^2
\end{equation}

where $\omega_a$ is the Kahler form corresponding to the complex structure
$I_a$, and we have used the fact that the value of the Casimir operator
$\sum_a I_a^2$ in the vector representation is $-8$. 

Consider the membrane, holomorphic in the complex structure $Je^{-I\phi}$,
where $\phi$ is some constant.
The restriction on this membrane of the form
\begin{equation}
\omega_{\phi}=
i\omega_I+\mbox{Im}\;(e^{-i\phi}\omega_C)
\end{equation}

is zero. Indeed, for such a membrane we may choose the vectors $\xi$
and $Je^{-I\phi}\xi$ as the basis of the tangent space, and
\begin{equation}\label{lagr}
\omega_{\phi}(\xi,Je^{-I\phi}\xi)=i\omega_{\phi}(\xi,\xi)=0
\end{equation}

It follows from (\ref{ds2}) and (\ref{lagr}), that 
the area of the holomorphic membrane is equal to
\begin{equation}
\int_M\mbox{Re}\;(e^{-i\phi}\omega_C)=
\left|\int_{\partial M}d^{-1}\omega_C\right|
\end{equation}

Thus, to find the masses of the BPS states,
we have to know an explicit expression
for $\omega_C$. We will obtain it as the Hamiltonian
reduction of the flat $\omega_C$ to the manifold 
$\mu_R=\mu_C=0$. On the flat space,
\begin{equation}
\omega_C=dy\wedge dz+ du\wedge dv
\end{equation}

Since on the constraint manifold $\mu_C=e$, $dv=-i(ydz+zdy)$, we get
\begin{equation}
\omega_C|_{\mu_C=e}=\left(du+i{dy\over y}\right)\wedge dv
\end{equation}

and we kill the $G_C$-symmetry by introducing $Y=e^{-iu}y$ to get
\begin{equation}
\omega_C=i{dY\over Y}\wedge dv
\end{equation}

The integral of this 2-form over the surface of the 2-brane
is reduced to the integral of $v dY/Y$ over the boundary. 
The new feature in the case of Taub-NUT compared to
$\R^3\times S^1$ is the existence of the membranes
whose boundary corresponds to odd electric charge. In the case
when the membrane goes through the singularity ($v=e$), the boundary
includes the small circle around $v=e$, because at this point $Y=0$,
and the differential has a singularity with the residue
\begin{equation}
\mbox{res}_{Y=0} vdY/Y=e
\end{equation}

Thus, the mass of the BPS state is given by
\begin{equation}
m=|iSe+n_e a +n_m a_D|
\end{equation}

where $S$ is the ``winding number'', which for the state corresponding
to the cycle $\Gamma$ is equal to $\int_{\Gamma}dY/Y$.

\section{Deformations of Fivebranes and Membranes.}\label{Deformations}

Here we will follow the method developed in \cite{DT} to study
the deformations of fivebrane and membranes.
Consider the complex submanifold 
$M\stackrel{i}{\subset} X$ of the HyperKahler manifold $X$.
We can describe the deformation of this manifold by the vector field
\begin{equation}
\xi\in\Gamma({\cal N}M)
\end{equation}

where ${\cal N}M$ is the normal bundle of the manifold $M$, whose
fiber at point $m\in M$ is 
${\cal N}_mM={\cal T}_{i(m)}X/i_*{\cal T}_mM$, $\Gamma$ denotes
the sections of the bundle over $M$. The vectors corresponding to complex
deformations (that is, the deformed surface remains complex) correspond
to the holomorphic sections:
\begin{equation}
\xi\in H^0(M,{\cal N}M)
\end{equation}

To study these holomorphic sections, we will follow the method
of \cite{DT}. Consider the map:
\begin{equation}
p:{\cal N}M \tilde{\rightarrow} {\cal T}^*M 
\end{equation}

given by the contraction of $\xi$ with the holomorphic symplectic
form, and then restricting to ${\cal T}M$:
\begin{equation}\label{map}
\xi\stackrel{p}{\mapsto} \iota_{\xi}\omega_C|_{{\cal T}M}
\end{equation}

Here $\omega_C$ is the holomorphic symplectic form, corresponding
to the complex structure of $M$.
In this formula we use any representative of  $\xi$ in ${\cal T}X$. All these
representatives differ by the vectors, tangent to $M$, and $\omega_C$
is zero on $M$. Thus, the right hand side of (\ref{map}) does
not depend on the choice of representative, and the map is correctly
defined. This is an isomorphism, since $M$ is a Lagrangian
manifold (that is, a maximal manifold, on which $\omega_C$ is zero). 

Thus, the deformations of complex submanifolds correspond to the holomorphic 
forms on them. There is a natural metric on the space of 
deformations:
\begin{equation}\label{KE}
||\xi||^2\stackrel{def}{=}\int_M d^2\sigma ||\xi_{\perp}||^2
\end{equation}

where $\xi_{\perp}$ is defined as the representative of $\xi$, orthogonal
to $M$, and $d^2\sigma$ is the surface element on $M$.
This metric can be rewritten as:
\begin{equation}\label{KE1}
||\xi||^2={1\over 2}\int_M \overline{p(\xi)}\wedge *p(\xi)
\end{equation}

Indeed, 
\begin{equation}
||\xi_{\perp}(m)||^2={1\over 2}||(J+iK)\xi_{\perp}(m)||^2=
{1\over 2}||\iota_{\xi_{\perp}}\omega_C||^2
\end{equation}

For any $\nu\in{\cal T}_m^*X$, 
$||\nu||^2=||\nu|_{{\cal T}_mM}||^2+
||\nu|_{\left({\cal T}_mM\right)^{\perp}}||^2$. But the restriction
of $\iota_{\xi_{\perp}}\omega_C$ on ${\cal T}^{\perp}M$ is zero, since
${\cal T}^{\perp}M$ is Lagrangian plane in ${\cal T}X$. Thus, 
${1\over 2}||\iota_{\xi_{\perp}}\omega_C||^2$ is equal to 
${1\over 2}||\iota_{\xi}\omega_C|_{{\cal T}M}||^2={1\over 2}||p(\xi)||^2$.

As an application of this formalism, consider first the possible
deformations of the curve $\Sigma$, representing the fivebrane.
The deformations, which have finite norm (\ref{KE}), correspond via
the map (\ref{map}) to the square-integrable meromorphic forms,
which is the same as holomorphic forms (decreasing at infinity of $\Sigma$),
or just square integrable harmonic forms. 
There are $2g$ of them, where $g$ is the genus of the (compactified) 
curve\footnote{As we explained in Section \ref{examples}, the
curve $\Sigma$ is topologically the genus $g$ surface $\bar{\Sigma}$ 
with some $n$ points omitted. The group 
$H_1^{comp}(\Sigma)=H_1(\bar{\Sigma})\oplus {\bf R}^{n-1}$ 
has thus dimension $2g+n-1$. But the cohomology class, whose pairing with
the cycles from $H_1(\bar{\Sigma})$ is zero, does not have
the square-integrable harmonic representative. Indeed, there are only 
$2g$ harmonic differentials on $\bar{\Sigma}$, and they correspond
to $2g$ cycles in $H_1(\bar{\Sigma})$. 
As was discussed in \cite{Witten}, the harmonic differential
corresponding to the cycle around the point at infinity 
necessarily has a pole at that point.}.
The integral (\ref{KE1}) may be rewritten using Stokes theorem:
\begin{equation}
||\xi||^2=\sum\limits_{ij}\Omega^{ij}
\int_{c^i}p(\xi)\int_{c^j}\overline{p(\xi)}
\end{equation}

where $\Omega^{ij}$ is the intersection pairing in $H_1(\Sigma,Z)$.
Since $\iota_{\xi}\omega$ is holomorphic, it represents the section
of the cotangent bundle to the $\mbox{Jac}(\Sigma)$, and the integral
can be written as
\begin{equation}\label{metric}
\int_{Jac \Sigma}\iota_{\xi}\omega\wedge
\overline{\iota_{\xi}\omega}\wedge t^{r-1}
\end{equation}

where $t$ is the polarization on $\mbox{Jac}\Sigma$. The massless vector
fields in the four-dimensional theory are also described in terms of
harmonic forms on $\Sigma$ \cite{Witten}. The metric (\ref{metric})
on the moduli space of five-brane agrees with the coefficients of the
kinetic terms of these vector fields, as it should be in supersymmetric
theory.

The relation between the holomorphic symplectic form
$\omega=ds\wedge dv$ and the Witten-Seiberg differential $\lambda$ is the
following. The meromorphic differential $\lambda$ is defined as
\begin{equation}
d\lambda=\omega
\end{equation}

We may take $\lambda=vds$. This differential is meromorphic,
but its derivative with respect to the moduli is the holomorphic
differential on the curve, up to maybe the derivative of the
meromorphic function:
\begin{equation}
\nu_{hol.}=\iota_{\xi}\omega=\iota_{\xi}d\lambda=-d\iota_{\xi}\lambda+
{\cal L}_{\xi}\lambda
\end{equation}

Here ${\cal L}_{\xi}\lambda$ may be thought of as the derivative 
with respect to the moduli.

Consider the deformations of the membranes. Again, the holomorphic
deformation vectors $\xi$ are related to holomorphic 1-forms $p(\xi)$ via
$\omega_C$. (If the membrane is holomorphic in the complex structure
$J$, we should take $\omega_C=\omega_K-i\omega_I$.)
Since the membrane has the boundary,
which is supposed to lie on the surface of the 5-brane, the appropriate
boundary conditions should be imposed on $\xi$. These boundary conditions
correspond to the following boundary conditions on $p(\xi)$.
Consider the harmonic 1-form 
\begin{equation}
h(\xi)=p(\xi)+\overline{p(\xi)}
\end{equation} 

The restriction of $h(\xi)$ on the boundary of the membrane should be zero.
Indeed, introduce the vector
$\eta$ directed along the intersection line of the membrane and the
fivebrane. The tangent planes to the 5-brane and membrane will be
$(\eta,I\eta)$ and $(\eta,J\eta)$. Notice that 
$\iota_{\xi_{\perp}}\omega_K ={1\over 2}h(\xi)$.  Thus, 
$\xi^a_{\perp}={1\over 2}K^a_bg^{bc}h_c$ is parallel to 
$KJ\eta=I\eta$, that is, belongs to the tangent space to $\Sigma$. 

The number of the harmonic forms with these boundary conditions
is equal to the first Betti number of the surface. 
For the genus $g$ surface with $n$ holes there are $2g+n-1$ of them. 
For the 
cylinder, there is one harmonic differential, $h=d\tau$, where
$(\phi,\tau)$ are the coordinates on the cylinder. This harmonic
differential corresponds to the zero mode, described in Section 
\ref{Complexu}.

\section{Membrane worldsheet theory.}
In this section we will explain, from the point of view of the
membrane worldsheet theory, why the type of the supermultiplet
depends on the topology of the membrane. 
The fields of the membrane worldsheet theory
\cite{BST,DWHN}
are 11 bosons
$X^{\mu}$ and their superpartners, the components of the
$SO(1,10)$ Majorana fermions $\Theta$. 
Let us introduce the light-cone coordinates
\begin{equation}
X^{\pm}={X^0\pm X^3\over\sqrt{2}}
\end{equation}

and choose the light-cone gauge
\begin{equation}
\begin{array}{c}
X^+(\zeta)=X^+(0)+\tau,\\
\gamma_+\Theta=0
\end{array}
\end{equation}

We have chosen one of the coordinates parallel to the 5-brane
worldvolume, $x^3$, as the longitudinal coordinate.

The reality condition for fermions is

\begin{equation}\label{reality}
\Theta^*={\cal C}\Theta
\end{equation}

where $\cal C$ is the $SO(9)$ charge conjugation matrix, characterized
by the property
\begin{equation}\label{CGC}
{\cal C}^{\dagger} \Gamma^A{\cal C}=(\Gamma^A)^T
\end{equation}

Let us choose the following representation for $SO(9)$ gamma-matrices:
\begin{equation}\label{DiracRep9E}
\begin{array}{ll}
\Gamma^{i+6}=\sigma^i\otimes\gamma^5\otimes\rho^3,  & i=1,2,3; \\
\Gamma^a=1\otimes \gamma^a\otimes\rho^3,  & a=4,5,6,10; \\
\Gamma^1=1\otimes 1\otimes\rho^1; &\\
\Gamma^2=1\otimes 1\otimes\rho^2&
\end{array}
\end{equation}

where $\sigma^i$ and $\rho^i$ are Pauli matrices, and $\gamma^a$ are
Euclidean gamma-matrices corresponding to $x^4,x^5,x^6,x^{10}$.

We will use the following charge conjugation matrix:
\begin{equation}\label{ccm}
{\cal C}=\sigma^2\otimes C\otimes \rho^1
\end{equation}

where $C$ is the four-dimensional Euclidean charge conjugation,
$C^{\dagger}\gamma^aC=-(\gamma^a)^T$.

For the membrane with the boundary on
the fivebrane the following boundary conditions should be imposed on
fermions:
\begin{equation}\label{bc}
{1\over 2}\left(1+\prod\limits_{j\in {\cal N}}\Gamma^j\right)S=0
\end{equation}

where $\cal N$ is the set of indices $j$ for which the boundary conditions
on $X^j$ are of Neumann type. The origin of these boundary conditions
is explained in \cite{BB}.
In our situation, the 5-brane is not flat, thus the boundary conditions
for fermions are not constant. The directions parallel to the 5-brane
world-volume, are spatial directions $x^1$ and $x^2$, the vector parallel
to the boundary of the membrane, and the rotation of this
vector by the operator $I$ (the complex structure of the 5-brane). Thus,
we have the following boundary conditions:

\begin{equation}\label{bks}
{1\over 2}\left(1+\Gamma^1\Gamma^2\Gamma^{\|}\right)S=0
\end{equation}

where

\begin{equation}
\begin{array}{c}
\Gamma^{\|}=1\otimes \gamma^{\|}\otimes 1\\
\gamma^{\|}=\gamma^{\partial_{\varphi}}\gamma^{I\partial_{\varphi}}\\
\gamma^{\partial_{\varphi}}={1\over \sqrt{|\dot{x}|^2+|\dot{y}|^2}}
(\dot{x}\gamma^x+\dot{y}\gamma^y+c.c.)\\
\gamma^{I\partial_{\varphi}}={i\over \sqrt{|\dot{x}|^2+|\dot{y}|^2}}
(\dot{y}^*\gamma^x-\dot{x}^*\gamma^y-c.c.)
\end{array}
\end{equation}

The group $SO(4)$, acting on the tangent space to the 
four-manifold, can be decomposed as the product of two $SU(2)$ groups:
\begin{equation}
SO(4)=(SU(2)_L\times SU(2)_R)/{\bf Z}_2
\end{equation}

where $SU(2)_L$ and $SU(2)_R$ act on the components of positive
and negative chirality. For the Hyper-Kahlerian manifold, the holonomy
group is $SU(2)_R$, and the group $SU(2)_L$ is generated by $I$, $J$
and $K$ --- the three complex structures. We will need the following
identity:
\begin{equation}\label{GaGb}
\gamma_a\gamma_b={1\over 2}\left(g_{ab}+
\sum\limits_n\omega_{ab}^n\hat{I}_n\right)
\end{equation}

(Here $\hat{I}_1$, $\hat{I}_2$ and $\hat{I}_3$ are the gamma-matrix
realizations of the complex structures, and $\omega_1$, $\omega_2$,
$\omega_3$ are the corresponding Kahler forms.) Indeed, for any two
vectors $\xi$, $\eta$, we have:
\begin{equation}
\hat{\xi}\hat{\eta}={1\over 2}
\left(g(\xi,\eta){\bf 1}+
{[}\hat{\xi},\hat{\eta}{]}\right)
\end{equation}

and the chiral part of the commutator can be represented as:
\begin{equation}
\begin{array}{c}
{[}\hat{\xi},\hat{\eta}{]}_L=
-{1\over 2}\sum\limits_n 
\mbox{tr}_L\left({[}\hat{\xi},\hat{\eta}{]}I_n\right)I_n
=\\=
{1\over 2}\sum\limits_n\mbox{tr}_L\left(\hat{\xi}\widehat{I_n\eta}
\right)I_n=\sum\limits_n\omega^n(\xi,\eta)I_n
\end{array}
\end{equation}

This identity implies that the $SU(2)_L$ component of $\gamma^{\|}$
is $\gamma_L^{\|}=\hat{I}$ (the gamma-matrix realization
of the 5-brane complex structure). Notice that the $SU(2)_R$ component
of $\gamma^{\|}$ is non-constant (depends on the point on the 
boundary). Thus, the boundary conditions for chiral fermions are:
\begin{equation}\label{bkschir}
\Gamma^1\Gamma^2S=\hat{I}S
\end{equation}

Fermionic zero modes are the solutions of the equations of motion
for fermions, which do not depend on the worldsheet time coordinate.
They satisfy the following
equation \cite{BST,CF}:
\begin{equation}\label{nablaS}
dX^a\Gamma_a\wedge\nabla S=0 
\end{equation}

where $\nabla$ is the pullback of the spin connection on the
membrane worldsurface. Since the holonomy group is $SU(2)_R$, the
positive chirality fermions are in trivial bundle. Thus, the
constant spinors of positive chirality are solutions of (\ref{nablaS}).
Let us show, that those constant spinors of positive chirality,
which satisfy the boundary condition (\ref{bkschir}), 
correspond to spatial spinors. 

Let us choose the tetrad $(e_x, e_y, e_{\bar{x}}, e_{\bar{y}})$
in such a way, that $SU(2)_L$ acts on the
coordinates of a vector as:
\begin{equation}
\begin{array}{c}
I\xi^x=i\xi^x,\;\;I\xi^y=i\xi^y\\
J\xi^x=i\overline{\xi^y},\;\; J\xi^y=-i\overline{\xi^x}
\end{array}
\end{equation}

One can check that the only metric with the property
$g(\xi,\eta)=g(I\xi,I\eta)=g(J\xi,J\eta)$ is
\begin{equation}
||\xi||^2=G(|\xi^x|^2+|\xi^y|^2)
\end{equation}

which means that the tetrad is orthogonal.

The groups $SU(2)_L$ and $SU(2)_R$ act on spinors as follows:
\begin{equation}
\begin{array}{lcc}
 & SU(2)_L & SU(2)_R \\
E& \gamma_y\gamma_x & \gamma_{\bar{x}}\gamma_y \\
H\;\;\;& 
  \left[\gamma_{x},\gamma_{\bar{x}}\right]+
  \left[\gamma_{y},\gamma_{\bar{y}}\right] \;\;\;&
  \left[\gamma_{\bar{x}},\gamma_x\right]-
  \left[\gamma_{\bar{y}},\gamma_y\right] \\
F& \gamma_{\bar{x}}\gamma_{\bar{y}} & \gamma_{\bar{y}}\gamma_x
\end{array}
\end{equation}

Spinors of positive chirality  
have the following form\footnote{To
satisfy equation (\ref{nablaS}), we could take arbitrary 
holomorphic $\phi_+^{(0)}(z)$ and antiholomorphic $\phi_-^{(0)}(\bar{z})$,
but only the constant $\phi_{\pm}$ can satisfy the boundary conditions.}:
\begin{equation}
\begin{array}{c}
S^{(0)}=
\gamma_{\bar{x}}\gamma_{\bar{y}}\phi_+^{(0)}+
\gamma_x\gamma_y\phi_-^{(0)}
\end{array}
\end{equation}

Substituting these solutions in the boundary conditions (\ref{bkschir})
we get:
\begin{equation}
\begin{array}{c}
\phi_-^{(0)}=
-\rho^3\gamma_{\bar{x}}\gamma_{\bar{y}}\phi_+^{(0)}
\end{array}
\end{equation}

Let us represent $\gamma_x$ and $\gamma_y$ in the space
${\bf C}^2\otimes {\bf C}^2$ as follows:
\begin{equation}\label{DiracRep4E}
\gamma_{x}=\left[\begin{array}{cc}0&0\\1&0\end{array}\right]\otimes 1,\;\;
\gamma_{y}=\left[\begin{array}{cc}1&0\\0&-1\end{array}\right]
\otimes\left[\begin{array}{cc}0&0\\1&0\end{array}\right],
 \;\;
\gamma_{\bar{x}}=\gamma_x^T,\;\; \gamma_{\bar{y}}=\gamma_y^T
\end{equation}

In this representation, the zero mode $S^{(0)}$ 
has the following form:
\begin{equation}\label{explicit}
S^{(0)}=\left[\begin{array}{c}a\\ 0 \\ 0 \\ -\rho^3  a
\end{array}\right]
\end{equation}

where $a$ and $b$ are arbitrary constant spinors of
$SO(3)_{\{7,8,9\}}\times SO(2)_{\{1,2\}}$.

Let us impose the reality conditions (\ref{reality}). For the
representation (\ref{DiracRep4E}), $C=i\tau^2\otimes\tau^1$,
thus the charge conjugation matrix (\ref{ccm}) is:
\begin{equation}
{\cal C}=\sigma^2\otimes (i\tau^2\otimes\tau^1)\otimes\rho^1
\end{equation}

and (\ref{reality}) gives the $SO(3)_{7,8,9}$ -- invariant conditions
\begin{equation}
a^*=i(\sigma^2\otimes \rho^2) a 
\end{equation}

which enable us to express the negative $SO(2)_{7,8}$-helicity component
of $a$ through the positive helicity component:
\begin{equation}
a=\left[\begin{array}{c}a_{\uparrow}\\
              \rho^2 a_{\uparrow}^*\end{array}\right]
\end{equation}

This means, that the zero modes are parametrized by
the spinors $a_{\uparrow}$ of the spatial $SO(3)$ group
(which is broken to $SO(2)$ by the gauge choice).

It turns out, that for the membranes with the topology
of the disk these constant $S$ are the only solutions of
the equations ({\ref{nablaS}), while for the surfaces with
more complicated topology, non-constant solutions
with negative chirality are possible.

These non-constant solutions are related to the harmonic
1-differentials on the membrane worldsuface, satisfying
the boundary conditions described in the end of 
Section \ref{Deformations}. For any such 
differential $h$, consider the vector field 
\begin{equation}\label{xiviah}
\xi^i=g^{ij}h_j
\end{equation}

where $g_{ij}$ is an induced metric on the membrane worldsurface.
Consider the operator 
\begin{equation}\label{Bh}
B_h=(i_*\xi)^a\Gamma_a
\end{equation}

acting on spinors. Here $i_*\xi$ is the vector $\xi$ on the
membrane surface, considered as the vector in the ambient space.
This operator has the following properties: it maps solutions
of (\ref{nablaS}) with positive chirality to solutions with
negative chirality, and also preserves boundary conditions and
the reality condition. Indeed, the vector $i_*\xi$ is orthogonal
to the tangent space of the 5-brane, thus $B$ respects boundary
conditions. The reality condition for $B_hS$ follows from 
(\ref{CGC}).
Let us check that $B_hS$ is the solution to (\ref{nablaS}):
\begin{equation}\label{nablaSB}
\begin{array}{c}
dX^a\Gamma_a \wedge \nabla\left((i_*\xi)^b\Gamma_b)S\right)=
-\nabla\left(dX^a\Gamma_a(i_*\xi)^b\Gamma_b S\right)=\\=
-{1\over 2}\nabla\left(dX^a g_{ab} (i_*\xi)^b S+
\sum\limits_n dX^a \omega^n_{ab}(i_*\xi)^b \hat{I}_n S\right)
\end{array}
\end{equation}

(Here $\hat{I}_1$, $\hat{I}_2$, $\hat{I}_3$ are the gamma-matrix 
realizations of the complex structures, and 
$\omega^1$, $\omega^2$, $\omega^3$ are the corresponding Kahler forms.)
The last row in (\ref{nablaSB}) is 
\begin{equation}
-{1\over 2}\nabla\left((h+*h\hat{J})S\right)=0
\end{equation}

--- this is zero, since both $h$ and $*h$ are closed forms, 
and $S$ is covariantly constant.

In the light-cone gauge, the supersymmetry on the membrane
worldsheet is generated by $\alpha$- and $\beta$-
transformations \cite{BST,DWHN}:
\begin{equation}\label{alphabet}
\begin{array}{c}
\delta X^I=2i\alpha^{\dagger}\Gamma^IS+
2i\epsilon^{ab}\partial_a X^I\alpha^{\dagger}\int d\tau
\partial_b S\\
\delta S=-{1\over 2\sqrt{2}P^{+}}\left(\dot{X}^I\Gamma_I
-{1\over 2}\epsilon^{ab}
\partial_aX^I\partial_bX^J\Gamma_{IJ}\right)\alpha+
2i\epsilon^{ab}\partial_aS\alpha^{\dagger}\int d\tau\partial_b S
+\beta
\end{array}
\end{equation}

Here $\alpha$ and $\beta$ are $SO(9)$  Majorana spinors, they have
together $32$ real components. Consider the membrane, holomorphic
in the appropriate complex structure. It is natural to consider
separately $\alpha$ and $\beta$ of positive and negative
chirality with respect to the group $SO(4)$ rotating 
$x^4,x^5,x^6,x^{10}$. On the non-flat Hyper-Kahlerian manifold,
only $\alpha$ and $\beta$ of positive chirality
generate rigid supersymmetry transformations.  
Even if the space is flat, those $\alpha$ and $\beta$ which 
have negative $SO(4)$ chirality give $\delta S$ which does
not satisfy the boundary conditions. Of those $\alpha$ and $\beta$
which have positive $SO(4)$ chirality, half give $\delta S$
satisfying the boundary conditions. Thus, we get $8$ of
$32$ supercharges preserved by the 5-brane. The BPS state,
corresponding to the membrane, should be annihilated by 
$4$ of these $8$ supercharges. 

Notice that the terms with $\partial S$ in (\ref{alphabet}) 
are zero for $\alpha$ with positive $SO(4)$ chirality. 
Consider the variation $\delta S$. It is zero if
 
\begin{equation}\label{betavsalpha}
\beta=-{1\over 2\sqrt{2}P^{+}}
\epsilon^{ab}\partial_a X^I\partial_b X^J \gamma_{IJ}\alpha
\end{equation}

For the holomorphic membrane,
\begin{equation}
{1\over P^+}\partial_a X^I\partial_b X^J \Gamma_{IJ}=
{1\over P_0^+}\hat{J}+\mbox{terms from $SU(2)_R$}
\end{equation}

where $\hat{J}$ is the gamma-matrix realization of the complex 
structure of the membrane, and the terms from $SU(2)_R$ are 
non-constant.

Then, the transformation $\delta_{\alpha}S$ with $\alpha$ of positive 
chirality can be compensated by constant $\beta$ given 
by (\ref{betavsalpha}).

Consider now the variation $\delta X^a$. For the (constant)
fermionic modes $S^{(0)}$ of positive chirality, the first formula
in (\ref{alphabet}) gives $\delta X^a$ corresponding to
the spatial translations (in the light cone gauge, there are
two transverse directions $x^1$ and $x^2$). Zero modes 
$B_hS^{(0)}$ of negative chirality give bosonic zero modes,
described in the previous section, and reparametrizations of the
membrane worldsheet:
\begin{equation}
\begin{array}{c}
\delta X^a=\alpha^{\dagger}\Gamma^a(i_*\xi)^b\Gamma_bS^{(0)}
=\\=
{1\over 2}\left( (i_*\xi)^a (\alpha^{\dagger}S^{(0)})+
\sum\limits_n (I_n i_*\xi)^a 
(\alpha^{\dagger}\hat{I}_nS^{(0)})\right)
\end{array}
\end{equation}

The terms with $(I i_*\xi)^a$ and $(i_*\xi)^a$ are zero. Indeed,
$S^{(0)}$ are can be written as $S^{(0)}=\hat{J}\alpha'$, where 
$\hat{I}\alpha'=-\Gamma_1\Gamma_2\alpha'$. Thus, 
\begin{equation}
\begin{array}{c}
\alpha^{\dagger}\hat{I}S^{(0)}=
\alpha^{\dagger}\hat{I}\hat{J}\alpha'=
-\alpha^{\dagger}\Gamma_1\Gamma_2\hat{J}\alpha'
=\\=
\alpha^{\dagger}\hat{J}\hat{I}\alpha'=
\alpha^{\dagger}\hat{J}\Gamma_1\Gamma_2\alpha'=0
\end{array}
\end{equation}

and for the same reason $\alpha^{\dagger}S^{(0)}=0$.
Thus, the only remaining transformations are
\begin{equation}
\delta X^a= (K i_*\xi)^a
\end{equation}

and
\begin{equation}
\delta X^a= (J i_*\xi)^a
\end{equation}

The first one is of the type described in section \ref{Complexu} --
the boundary of the membrane slides along the 5-brane. 
The second is an infinitesimal reparametrization of the membrane
worldsheet. 

For the disk, quantization of fermionic zero modes gives four
bosonic states
\begin{equation}
|0>,\;\; a_1a_2|0>,\;\; |\tilde{0}>,\;\; a_1 a_2|\tilde{0}>
\end{equation}

and four fermionic states
\begin{equation}
a_1|0>,\;\; a_2|0>,\;\; a_1|\tilde{0}>,\;\; a_2|\tilde{0}>
\end{equation}

where $|0>$ and  $|\tilde{0}>$ correspond to the disks of the opposite
orientation (monopole and antimonopole).  This is the hypermultiplet.
Notice, that this is indeed a BPS state. We can make $\delta S=0$ by
choosing $\beta$ as in (\ref{betavsalpha}). Then, the first equation
in (\ref{alphabet}) gives us $\delta X$ corresponding to spatial 
translations, and we require the wave function to be constant over
$x^1$, $x^2$ (notice that we are working in the centre of mass frame).

For the cylinder, we have also non-constant modes of negative chirality.
We will call them $b$-modes. The supersymmetry transformations
of these $b$-modes give the bosonic zero mode, corresponding
to the moduli space of cylinders.
The BPS states correspond to the vacua of 
supersymmetric quantum mechanics on this moduli space 
\cite{PhaseTransitions}. There are four supercharges
\begin{equation}\label{supercharges}
b_1{\partial\over\partial m},\;\;
b_2{\partial\over\partial m},\;\;
b_1^*{\partial\over\partial m},\;\;
b_2^*{\partial\over\partial m}
\end{equation}

which should annihilate the wave function. We have denoted the
collective coordinate $m$. 
The Hamiltonian is ${\partial^2\over\partial m^2}$. 
Since the moduli space is the interval $\left[m_l,m_r\right]$,
some boundary conditions should be imposed (otherwise the Hamiltonian
is not self-conjugate).
As we explain in Appendix B, for the wave function
\begin{equation}
\psi(m)=\psi_0(m)+b_1\psi_1(m)+b_2\psi_2(m)+b_1b_2\psi_{12}(m)
\end{equation}

the only possible supersymmetric boundary conditions are either
\begin{equation}\label{1}
\begin{array}{c}
\partial_m \psi_0=\partial_m \psi_{12}=0 \\
\psi_1=\psi_2=0
\end{array}
\end{equation}

or
\begin{equation}\label{2}
\begin{array}{c}
\partial_m\psi_1=\partial_m\psi_2=0\\
\psi_0=\psi_{12}=0
\end{array}
\end{equation}

--- these equalities should be satisfied at the ends of the
interval, $m=m_l$ and $m=m_r$.

Since the vacuum wave function should be 
constant, the boundary conditions of the first type leave us 
with the states
\begin{equation}
\psi_0+b_1b_2\psi_{12}
\end{equation}

which have spin zero, and the second type gives
\begin{equation}
b_1\psi_1+b_2\psi_2
\end{equation}

which have spin $1/2$. 
We should also act on these states by the $a$-modes.
This gives us the hypermultiplet for the boundary
conditions of the type (\ref{1}), and the vector multiplet
for the boundary conditions of the type (\ref{2}).

Understanding which type of boundary conditions should be
imposed probably requires considering the theory at short
distances. 
There are at least two cases when it may be important to know
what happens at short distances. The first case is
when the modulus of  $\Sigma$ intersects the curve of marginal 
stability. The boundary of the cylinder touches itself and the 
cylinder decays into two disks. These disks do not have $b$-modes,
only the constant $a$-modes. We cannot explain how $b$-mode for the
cylinder transforms into the $a$-mode for the disk in the low energy
theory. 
It follows from expression (\ref{xiviah}), (\ref{Bh}),
that when the length of the cylinder goes to zero,
the $b$-modes become localized in the very small region near the
points where the boundary touches 
itself\footnote{Explicit expressions for the $b$-modes contain
terms like ${\partial x\over |\partial x|^2+|\partial y|^2}$,
${\partial y\over |\partial x|^2+|\partial y|^2}$ --- 
cf. (\ref{cyl}). }. 
Indeed, 
when the length of the cylinder is very small,
$dx$ and $dy$ should be very large everywhere
except for that region. Thus, the complete understanding of what 
happens to the fermionic zero modes when the cylinder decays
requires considering the theory at short distances.   

The other case is when we consider the boundary conditions
for the wave function of the cylinder at the boundary of the moduli
space. As we have discussed in Section \ref{Complexu}, the 
boundary of the cylinder moduli space corresponds to
the cylinder degenerating into the disk. At this point,
the boundary touches itself and the same problem with the
$b$-modes appear. It should be true that the boundary conditions
of the type (\ref{2}) appear.

Notice that the $SO(3)_{7,8,9}$ group, realized by $a_1a_2+b_1b_2$,
$a_1^*a_2^*+b_1^*b_2^*$ and $\left[ a_1,a_1^*\right]+
\left[ a_2,a_2^*\right]+\left[b_1,b_1^*\right]+
\left[b_2,b_2^*\right]$, acts on the states in four-dimensional
theory as $SU(2)_{\cal R}$ symmetry.

\section{Membranes of minimal area and geodesics.}
In this section we will show that the boundaries of the minimal
surfaces are not necessarily geodesics. We will explain
how our discussion of BPS states is related to the one 
in \cite{KLMVW}. We will first consider two examples. 

The simplest example arises in the theories with matter in
mixed representation.
Let us consider the ``spike'':
\begin{equation}
v=m/s
\end{equation}

In the case of the pure spike (without regular terms)
the boundary of the membrane is geodesic. Indeed, it is described
by the equation $|s|^2=m$, and the meromorphic
differential $$\lambda=vds=mds/s$$ has constant phase (purely
imaginary) on any vector tangent to the boundary. This means,
that it is geodesic in the metric $|\lambda|^2$.

But, if we add regular terms, then the boundary is generally not
geodesic any more. For example, consider
\begin{equation}
v=m/s+\epsilon s
\end{equation}

where $\epsilon$ is very small. Then, the equation for the membrane
is
\begin{equation}
v-s^*=\epsilon{v^*+s\over 2}+o(\epsilon)
\end{equation}

and the boundary is
\begin{equation}
|s|^2=m+o(\epsilon)
\end{equation}

Now
\begin{equation}
vds=mi(1+\epsilon e^{2i\varphi}+o(\epsilon))d\varphi
\end{equation}

so the phase is changing, and the boundary is not geodesic.

As the second example we take the pure $SU(2)$.
Consider a very light monopole, that is $u$ close to $1$.
Put $x^6+ix^{10}=i\pi+s$. The monopole is located in the
region of small $s$ and $v$, where the curve may be represented
as
\begin{equation}\label{smallcurve}
-{s^2\over 2}-\epsilon s^4 -v^2+(\rho+i\sigma)+o(\epsilon)=0
\end{equation}

(after appropriate rescaling of $s$ and $v$ by the same factor.
Here $\rho+i\sigma$ is related to $u-1$ by the appropriate
rescaling.)

We consider the equation for the membrane to the first order in
$\epsilon$. Describe the membrane by the ansatz:
\begin{equation}
s^*-v=(-i+a)(s+v^*)+ib(s+v^*)^3
\end{equation}

where $a$ and $b$ are real constants of the order $\epsilon$.
We may rewrite it as
\begin{equation}\label{ansatz}
\begin{array}{c}
s=\left[ (i+a)+(3|v|^2-|s|^2)b\right]s^* \\
v=\left[ (i-a)+(3|s|^2-|v|^2)b\right]v^*
\end{array}
\end{equation}

To find the boundary, we have to use these equations together
with (\ref{smallcurve}). After substituting (\ref{ansatz})
in (\ref{smallcurve}) we get
\begin{equation}\label{im}
{1\over 2}|s|^2+|v|^2=\sigma
\end{equation}

for the imaginary part and
\begin{equation}\label{re}
\epsilon |s|^4+\rho -{1\over 2}\left[a+(3|v|^2-|s|^2)b\right]|s|^2
-\left[-a+(3|s|^2-|v|^2)b\right]|v|^2=0
\end{equation}

for the real part of (\ref{smallcurve}). The condition that
the membrane intersects the curve by  the line means that
(\ref{re}) follows from (\ref{im}).
This requirement gives
\begin{equation}
\begin{array}{c}
a=-{11\over 2}b\sigma\\
b=-{1\over 3}\epsilon\\
\rho={9\over 2}b\sigma^2
\end{array}
\end{equation}

Actually we do not need the values of these parameters
in addressing the question of the phase of $vds$. Using
the equations
(\ref{im}) and (\ref{ansatz}), we can write an equation for
the boundary of the membrane, parametrized by the angle $\varphi$:
\begin{equation}
\begin{array}{c}
v=\sqrt{\sigma}\cos\varphi e^{i\pi\over 4}
\left(1+{ia\over 2}-(3|s|^2-|v|^2){ib\over 2}+o(\epsilon)\right)\\
s=\sqrt{2\sigma}\sin\varphi e^{i\pi\over 4}
\left(1-{ia\over 2}-(3|v|^2-|s|^2){ib\over 2}+o(\epsilon)\right)
\end{array}
\end{equation}

And for the differential we have:
\begin{equation}
\begin{array}{c}
v{ds\over d\varphi} =\sqrt{2}\sigma^2i\left[
\cos^2\varphi \left(1-2(|v|^2+|s|^2){ib\over 2}\right)
\right.
-\\-
\left.
\cos\varphi\sin\varphi{d\over d\varphi}(3|v|^2-|s|^2){ib\over 2}
\right]
=\\=
\sqrt{2}\sigma^2i\cos^2\varphi \left[1-ib\sigma^2(1-4\sin^2\varphi)
\right]
\end{array}
\end{equation}

The phase of this expression is not constant.

The boundaries of the membranes are more complicated then geodesics.
They depend essentially on the embedding of the curve into
$\mbox{\bf R}^3\times S^1$.
Consider two embeddings:
\begin{equation}\label{odin}
\cosh s+u-v^2=0
\end{equation}

and
\begin{equation}\label{dva}
\cosh s+u -\alpha^2 v^2=0
\end{equation}

where $\alpha$ is constant. These two curves are isomorphic as complex
curves, the isomorphism given by $v\to \alpha v$. And the geodesics of
the metrics $|vds|^2$ are preserved by this isomorphism. But this is
not true for the boundaries of the minimal surfaces: their dependence
on $\alpha$ is quite complicated. Indeed, the shape of the minimal surface
depends on the metric in the ambient space. The minimal surface 
for the curve (\ref{dva}) would be related to the minimal surface 
for (\ref{odin}), if the metric in the space $\R^3\times S^1$, where
(\ref{dva}) is embedded, were not $|ds|^2+|dv|^2$, but 
$|ds|^2+|\alpha dv|^2$.  Notice however
the simple relation between the masses
of the membranes with boundaries on (\ref{dva}) and the masses of 
the membranes ending on (\ref{odin}):
\begin{equation}
m_{\alpha}={1\over|\alpha|}m_1
\end{equation}

Let us prove that in the limit $\alpha\to\infty$ 
the boundaries
of the membranes become geodesics in the metric $|vds|^2$. This
means, that the value of the differential $vds$ on the vector, 
tangent to the boundary of the membrane, has constant phase:
\begin{equation}
\mbox{Arg}\left(v{ds\over dt}\right)=\mbox{const}
\end{equation}

We assume that $\alpha$ is real.
It is useful to introduce the coordinate $V=\alpha v$.
The equation for $\Sigma$ is:
\begin{equation}
\cosh s+P(V)=0
\end{equation}

Since the membrane is holomorphic in coordinates $(x,y)$, 
the restriction of the form
\begin{equation}\label{dxdy}
\begin{array}{c}
4dx\wedge dy=ds\wedge ds^*+d\tilde{v}
\wedge d\tilde{v}^*+ds\wedge d\tilde{v}+ds^*\wedge d\tilde{v}^*
=\\=
ds\wedge ds^* +{1\over\alpha^2}d\tilde{V}\wedge d\tilde{V}^*+
{1\over\alpha}\left(ds\wedge d\tilde{V}+ds^*\wedge d\tilde{V}^*\right)
\end{array}
\end{equation}

on the surface of the membrane iz zero. In the imaginary part of this 
expression, we may neglect the term 
${1\over\alpha^2}d\tilde{V}\wedge d\tilde{V}^*$, 
small compared to $ds\wedge ds^*$. 
This gives the condition:
\begin{equation}
ds\wedge ds^*|_{\Lambda^2{\cal T}M}=0
\end{equation}

meaning that the projection of the membrane on the surface $s$
has real dimension one. In other words, the membrane lies 
in three-dimensional surface $C\times \C_v$, 
where $C$ is some curve of real dimension one in the $s$-plane, 
and $\C_v$ is the $v$-plane. 

The real part of (\ref{dxdy}) is:
\begin{equation}
{1\over\alpha}\mbox{Re} (ds\wedge d\tilde{V})=0
\end{equation}

This implies, that the section of the membrane by the plane $s=s_0$
is a line interval. If the tangent vector to $C$ at point $s_0$ is
$\dot{s}$, then this line interval is given by the equation:
\begin{equation}
v=v_0+{i\over\dot{s}}\kappa
\end{equation}

where $\kappa$ is a (real) parameter along the line.
This gives the following description of the membrane boundary.
Consider two solutions $v_1(s)$ and $v_2(s)$ of
\begin{equation}
\cosh s+P(v)=0
\end{equation}

The tangent vector to the boundary of the membrane is
\begin{equation}
\dot{s}={e^{i\phi}\over v_1(s)-v_2(s)},\;\;\; \phi=\mbox{const}
\end{equation}

In particular, for the $SU(2)$ case, $v_2=-v_1=-v(s)$, and we get
\begin{equation}
v(s)\dot{s} = e^{i\phi}
\end{equation}

which means, that the boundary is a geodesic line.

\section{Appendix A. Complex structures of the Taub-NUT space.}

One of the complex structures is just
\begin{equation}
I.\left[
\begin{array}{c} \xi_Y \\ \xi_Z \\ \xi_v \end{array}
 \right] =
 \left[
\begin{array}{c} i\xi_Y \\ i\xi_Z \\ i\xi_v \end{array}
 \right]
\end{equation}

Let us calculate the other complex structure, $J$.
On the tangent, space to $\H\times \H$, the other complex structure
acts as the left multiplication on $j$:
$(\delta q,\delta s)\to (j\delta q, j\delta s)$, or:
\begin{equation}
J.\left[
\begin{array}{c}
\xi_y \\ \xi_z \\ \xi_u \\ \xi_v
\end{array}
\right]=
\left[
\begin{array}{c}
-\xi_z^* \\ \xi_y^* \\ -\xi_v^* \\ \xi_u^*
\end{array}
\right]
\end{equation}

Given the equivalence class $(Y,Z)$ of the points on the
manifold $\mu_C=0$ modulo $G_C$, we can always
find the corresponding point on $\mu_C=\mu_R=0$ (modulo $G$).
Let it be
\begin{equation}
Z=exp(iu)z;\;\; Y=\exp(-iu)y
\end{equation}

Then, to find $u_{im}=\mbox{Im}u$ we may use the equation
\begin{equation}\label{trans}
\mu_R={1\over 2}(\exp(-2u_{im})|Y|^2-\exp(2u_{im})|Z|^2-2u_{im})=0
\end{equation}

-- this is easy to see that this equation always has a
solution, and the solution is unique.

So, we have the vector $(\xi_Y,\xi_Z)$, tangent to $\mu_C=0$ at
the point $(Y,Z)$.
We construct the vector
$(\xi_y,\xi_z)=(e^{iu}\xi_Y, e^{-iu}\xi_Z)$,
at the point $(y,z)$ in $\mu_R=\mu_C=0$, tangent to $\mu_C=0$.
We now add to $(\xi_y,\xi_z)$
the vector of the form
$\left[ ity, -itz, t, 0 \right]$ so that
acting by $J$ on the resulting vector
we get again a vector tangent to $\mu_C=0$:
\begin{equation}
t^*=-i
\frac{i\xi_u^*-z\xi_z^*+y\xi_y^*}{1+|z|^2+|y|^2}
\end{equation}

Now, acting by $J$ and going to coordinates $Z$, $Y$, we get:
\begin{equation}\begin{array}{c}
\left[
\begin{array}{c} \xi_Y \\ \\ \xi_Z \end{array}
\right]
\stackrel{J}{\rightarrow}
i
\left[
\begin{array}{cc}
-{2+\rho\over 1+\rho}YZ^*&
-{2+|y|^{-2}+\rho\over 1+\rho}YY^*\\ \\
{2+|z|^{-2}+\rho\over 1+\rho}ZZ^* &
{2+\rho\over 1+\rho}ZY^*
\end{array}
\right]
\left[
\begin{array}{c}\xi^*_Y \\ \\ \xi^*_Z \end{array}
\right]
\end{array}
\end{equation}

where $\rho=|y|^2+|z|^2$.
Notice, that in this expression $|y|^2$ and $|z|^2$ are related to
$|Y|^2$ and $|Z|^2$ by multiplication on $\exp(2u_{im})$ and
$\exp(-2u_{im})$ correspondingly, where $u_{im}$ may be found from
the transcendent equation (\ref{trans}).

Now $K$ can be found from $K=JI$.
When $e\to\infty$ and $v$ remains finite,
both $Z$ and $Y$ are large.
We keep $u_{im}$ finite, so $|z|^2$ and $|y|^2$ are also large.
We find
\begin{equation}
J.\left[
\begin{array}{c}\xi_Y\\ \xi_Z\end{array}
\right]=\left[
\begin{array}{c}
-Y(Z^*\xi^*_Y+Y^*\xi^*_Z)\\ Z(Z^*\xi^*_Y+Y^*\xi^*_Z)
\end{array}\right]
\end{equation}

Using $\xi_v=i(Z\xi_Y+Y\xi_Z)$, and going to $s=\log Y$, we get
\begin{equation}
\begin{array}{c}
J.\xi_s=i\xi^*_v\\
J.\xi_v=-i\xi^*_s
\end{array}
\end{equation}

which is the same as the $J$ complex structure in the flat space.

\section{Appendix B. Supersymmetric quantum mechanics on the interval.} 

Consider the wave function of the form
\begin{equation}
\psi=\psi_0+b_1\psi_1+b_2\psi_2+b_1b_2\psi_{12}
\end{equation}

For the Hamiltonian to be self-conjugate, we have to require
for the boundary values:
\begin{equation}
\left[\begin{array}{c}\partial_m\psi\\ \psi\end{array}\right]
\in L
\end{equation}

where $L$ is the $\C$-linear subspace in ${\bf C}^8$, such that
for any two vectors 
$\left[\begin{array}{c}p_1\\q_1\end{array}\right]$ and
$\left[\begin{array}{c}p_2\\q_2\end{array}\right]$ in $L$:
\begin{equation}\label{Lagrangian}
(\overline{p_1},q_2)-(\overline{q_1},p_2)=0
\end{equation}

These boundary conditions should be supersymmetric. That is,
the space of functions such that the boundary values 
$\left[\begin{array}{c}\partial_m\psi\\ 
\psi\end{array}\right]\in L$ should
be preserved by 
\begin{equation}\label{four}
b_1\partial_m,\;\;b_2\partial_m,\;\;b_1^*\partial_m,\;\;
b_2^*\partial_m
\end{equation}

Our moduli space is an interval $m\in\left[m_l,m_r\right]$. 
Consider the functions, whose Fourier coefficients in the series
\begin{equation}
\psi(m)=\sum\limits_n
\left[a_n\cos \left({2\pi n\over m_l-m_r}m\right)+
b_n\sin \left({2\pi n\over m_l-m_r}m\right)\right]
\end{equation}

decrease sufficiently rapidly with $n$. Then,
\begin{equation}
{\partial^2\over \partial m^2}\psi(m)=
-\sum\limits_n \left({2\pi n\over m_l-m_r}\right)^2
\left[a_n\cos \left({2\pi n\over m_l-m_r}m\right)+ 
      b_n\sin \left({2\pi n\over m_l-m_r}m\right)\right]
\end{equation}

Let us describe all the possible subspaces in the space 
of boundary values, which are preserved by (\ref{four}).
We have to require that $L$ is preserved by the operators
\begin{equation}
\left[\begin{array}{cc}
0&-\left({2\pi n\over m_l-m_r}\right)^2 b_i\\
 b_i&0\end{array}\right]
,\;\;\;\;\;
\left[\begin{array}{cc}
0&-\left({2\pi n\over m_l-m_r}\right)^2 b_i^*\\
 b_i^*&0\end{array}\right]
\end{equation}

where $i=1,2$ and $n$ an arbitrary integer. This means, that if
\begin{equation}
\left[\begin{array}{c} p\\q\end{array}\right]\in L
\end{equation}

then $L$ contains also the vectors of the form
\begin{equation}
\left[\begin{array}{c}\sum\limits_i
(\alpha_ib_i+\beta_ib_i^*)q\\ 0\end{array}\right]
\;\;\;\mbox{and}\;\;\;
\left[\begin{array}{c}
0\\ \sum\limits_i(\alpha_ib_i+\beta_ib_i^*)p\end{array}\right]
\end{equation}

for arbitrary $(\alpha_i,\beta_i)$. A straightforward computation 
shows that if $v=v_0+b_1v_1+b_2v_2+b_1b_2v_{12}\neq 0$, then
\begin{equation}
\mbox{dim}_{\C}\left(\C b_1+\C b_2+\C b_1^*+\C b_2^*\right)v\geq 2
\end{equation}

with equality if and only if $v$ is an eigenvector
of $(-1)^F$, that is, $v$ is either of the form $b_1v_1+b_2v_2$,
or of the form $v_0+b_1b_2v_{12}$. Taking into account that 
the dimension should be $\mbox{dim}_{\bf C}L=4$, we get the only 
two possibilities:

1) $L$ consists of the vectors of the form
\begin{equation}
\left[\begin{array}{c} 
p_0+b_1b_2p_{12}\\
b_1q_1+b_2q_2\end{array}\right]
\end{equation}

2) $L$ consists of the vectors of the form
\begin{equation}
\left[\begin{array}{c}
b_1p_1+b_2p_2\\
q_0+b_1b_2q_{12}
\end{array}\right]
\end{equation}

In both cases, $L$ satisfies the property (\ref{Lagrangian}) .

\section{Acknowledgments.} I would like to thank Prof.~E.~Witten
for suggesting the problem and for advice. 
Discussions with S.~Gukov
and N.~Nekrasov were very helpful. 
This work was supported in part by RFFI Grant No. 96-02-19086.

\end{document}